\documentstyle[aps,epsf,epsfig]{revtex}

\newcommand{\bq}{\begin{equation}}
\newcommand{\eq}{\end{equation}}
\newcommand{\bqn}{\begin{eqnarray}}
\newcommand{\eqn}{\end{eqnarray}}
\newcommand{\nb}{\nonumber}
\newcommand{\lb}{\label}

\begin{document}
\title{Comment on ``Absence of trapped surfaces and singularities in
cylindrical collapse"}
\author{ Anzhong Wang  \thanks{ E-mail: Anzhong$\_$Wang@baylor.edu}}
\address{ CASPER, Physics Department, Baylor University,
Waco, TX76798-7316\\
and \\
Department of Physics, University of Illinois at
Urbana-Champaign, 1110 West Green Street, Urbana, IL61801-3080}
 
\date{\today}
\maketitle

\begin{abstract}

Recently, the gravitational collapse of an infinite cylindrical thin shell
of matter in an otherwise empty spacetime with two hypersurface orthogonal
Killing vectors was studied by  Gon\c{c}alves [Phys. Rev. {\bf D65}, 084045
(2002).]. By using three ``alternative" criteria for trapped surfaces, the
author claimed to have shown that {\em they can never form either outside
or on the shell, regardingless of the matter content for the shell, except
at asymptotical future null infinite}.

Following Penrose's original idea, we first define trapped surfaces in
cylindrical spacetimes in terms of the expansions of null directions orthogonal
to the surfaces, and then show that the first criterion used by Gon\c{c}alves 
is incorrect. We also show that his analysis of non-existence of trapped 
surfaces in vacuum is incomplete. To confirm our claim, we present an example 
that is a solution to the vacuum Einstein field equations and satisfies all the 
regular conditions imposed by Gon\c{c}alves. After extending
the solution to the whole spacetime,  we show explicitly that trapped surfaces  
exist in the extended region.

\end{abstract}

\vspace{.6cm}


\section{Introduction}
\lb{SecI}

The gravitational collapse of an infinite cylindrical thin shell
of generic matter in an otherwise empty spacetime with two hypersurface
orthogonal Killing vectors was recently studied by  Gon\c{c}alves
\cite{Gon02}. After re-deriving the junction conditions across an infinitely
thin shell with the spacetime being vacuum both inside and outside the shell,
the author studied the formation of trapped surfaces by  three ``alternative"
criteria, the so-called proper circumference radius criterion,  the specific area
radius criterion, and the criterion of geodesic null congruences. The author
claimed  that {\em trapped surfaces can never form either out of the shell or
on the shell, except at asymptotic future null infinity}.

In this Comment, following Penrose's original  idea \cite{Pen68}, we shall
first  define trapped surfaces in cylindrically symmetric
spacetimes in terms of the expansions of the two null directions orthogonal to the
surfaces. Then we shall show that the first criterion used by Gon\c{c}alves
for the formation of trapped surfaces 
is incorrect. We shall also show that Gon\c{c}alves' 
analysis of non-existence of
trapped surfaces in vacuum by the second and third criteria  is incomplete.
The reason is simply because the Einstein-Rosen coordinates used by Gon\c{c}alves
usually cover only  part of a spacetime, quite similar to the Schwarzschild
coordinates in the spherically symmetric spacetimes.  To show this explicitly, we
present  an example that is a solution to the vacuum Einstein field equations and
satisfies all the conditions that Gon\c{c}alves imposed for a spacetime to
be cylindrical. In the Einstein-Rosen coordinates,  trapped surfaces
indeed do not exist. But, after extending the solution to the whole spacetime, 
we do find trapped surfaces in the extended region.

\section{Trapped Spatial Two-Surfaces in Cylindrical Spacetimes}
\lb{SecII}

The general metric for cylindrical spacetimes with two hypersurface orthogonal Killing
vectors takes the form \cite{Kramer80},
\bq
\lb{2.1}
ds^{2} = e^{2(\gamma - \psi)}\left(dr^{2} - dt^{2}\right)
+ e^{2\psi}dz^{2} + \alpha^{2} e^{-2\psi}d\varphi^{2},
\eq
where $\gamma,\; \psi$ and $\alpha$ are functions of $t$ and $r$ only, and
$x^{\mu} = \{t, r, z, \varphi\}$ are the usual cylindrical coordinates,  and the
hypersurfaces $\varphi = 0, 2\pi$ are identified.
To have cylindrical symmetry, some   conditions needed to be
imposed. In general this is not trivial 
\cite{Barnes}.  Gon\c{c}alves defined cylindrical
spacetimes  by the existence of the two commuting spacelike Killing vector
fields, $\xi_{(2)} \equiv \partial_{z}$ and $\xi_{(3)} \equiv \partial_{\varphi}$.
At the symmetry axis $r = 0$,  the local-flatness condition was also imposed,
\bq
\lb{cd1}
\frac{X_{,\alpha}X_{,\beta} g^{\alpha\beta}}{4X} \rightarrow  1,
\eq
as  $r \rightarrow 0^{+}$, where $(\;)_{,\alpha} \equiv \partial (\;)/\partial
x^{\alpha}$, $ X \equiv \xi^{\mu}_{(3)}\xi^{\nu}_{(3)}g_{\mu\nu}$, and the radial
coordinate is chosen such that the symmetry axis is  at $r = 0$.

The concept of {\em trapped surfaces} was originally from Penrose \cite{Pen68},
who defined it as a  {\em compact} spatial two-surface $S$ on which
$\left.\theta_{+}\theta_{-}\right|_{S}
> 0$, where $\theta_{\pm}$ denote the expansions in the future-pointing null
directions orthogonal to $S$, and the spacetime is assumed to be time-orientable,
so that ``future" and ``past" can be assigned consistently. One may then define
a past trapped surface by $\left.\theta_{\pm}\right|_{S} > 0$, and a future
trapped surface by $\left.\theta_{\pm}\right|_{S} < 0$.

Recently, Hayward generalized the above definition to the cylindrical spacetimes
where the two-surface $S$ is {\em not compact} but  an infinitely long two-cylinder
of constant $t$ and $r$, and call it trapped, marginal or untrapped,
according to where $\alpha_{,\mu}$ is timelike, null or spacelike \cite{Hay00}.
This definition is the second criterion used in \cite{Gon02}, and was referred
to as the specific area radius criterion.

In the vacuum case,  one of the Einstein field equations can be written as
\cite{Kramer80}
\bq
\lb{2.15}
\alpha_{,tt} - \alpha_{,rr} = 0,
\eq
which has the general solution
\bq
\lb{2.16}
\alpha(t, r) =  F(t+r) - G(t-r),
\eq
where $F(t+r)$ and  $G(t-r)$ are arbitrary functions of their indicated arguments.
In the Einstein-Rosen coordinates, the functions $F$ and $G$ are chosen as
$F(t+r) = (t+r)/2$ and $G(t-r) = (t-r)/2$,  so that
\bq
\lb{2.16a}
\alpha = r.
\eq
Working with the Einstein-Rosen gauge,  Gon\c{c}alves calculated the expansions
in the future-pointing null directions orthogonal to a cylinder of constant $t$ and
$r$, and found that
\bq
\lb{2.2}
\theta_{\pm}^{Gon.}  = \left(\partial_{t} \mp \partial_{r}\right)
\left(r e^{-\psi}\right) = - re^{-\psi}\left(\left(\partial_{t}
\mp \partial_{r}\right)\psi \pm \frac{1}{r}\right).
\eq
These expressions are wrong and can be seen clearly from Eq.(B7) given in 
\cite{Wang03}, which in terms of the present notation takes the form 
\bqn
  \lb{2.7}
\theta_{+} &\equiv& g^{\alpha\beta} l_{\alpha;\beta}
   = e^{-2\sigma}\frac{\alpha_{,v}}{\alpha}
   = e^{2(\psi - \gamma)}
   \frac{\alpha_{,t} + \alpha_{,r}}{2b'(u)\alpha},\nb\\
\theta_{-} &\equiv& g^{\alpha\beta} n_{\alpha;\beta}
  = e^{-2\sigma}\frac{{\alpha}_{,u}}{\alpha}
  = e^{2(\psi - \gamma)}
   \frac{\alpha_{,t} - \alpha_{,r}}{2a'(v)\alpha},
\eqn
where 
\bq
\lb{2.3a}
t \equiv a(v) + b(u),\;\;\;
 r \equiv a(v) - b(u), 
\eq
with $a(v)$ and $b(u)$ being arbitrary 
functions of their indicated arguments, subject to $a'(v)b'(u) > 0$. A prime 
denotes the ordinary differentiation. 
The two null vectors $l_{\lambda}$ and $n_{\lambda}$ are future-directed, 
orthogonal to the two-cylinder of constant $t$ and $r$, and given by
\bqn
\lb{2.3b}
l_{\lambda} &\equiv&  
 \delta^{u}_{\lambda} = \frac{1}{2b'(u)}\left(\delta^{t}_{\lambda}
 - \delta^{r}_{\lambda}\right),\nb\\
 n_{\lambda} &\equiv&  
 \delta^{v}_{\lambda} = \frac{1}{2a'(v)}\left(\delta^{t}_{\lambda}
 + \delta^{r}_{\lambda}\right).
\eqn
For the details, we refer readers to \cite{Wang03}.
It should be noted that the two null vectors  
are uniquely defined only up to a factor \cite{HE73}. In fact, $\bar{l}_{\mu} 
= f(u)\delta^{u}_{\mu}$ and $\bar{n}_{\mu} = g(v)\delta^{v}_{\mu}$ represent another
set of null vectors that also define affinely parameterized null 
geodesics, and  the corresponding expansions
are given by $\bar{\theta}_{+} = f(u)\theta_{+}$ and $\bar{\theta}_{-} = g(v)\theta_{-}$. 
However, since along each curve $u = Const.$ ($v = Const.$) $f(u)$ ($g(v)$) is constant, 
this does not affect the definition of trapped surfaces in terms of the expansions (See 
\cite{HE73} and the discussions given below). Thus, without loss of generality,
in the following we consider only the expressions given by Eq.(\ref{2.7}).

Once we have $\theta_{\pm}$, following Penrose and Hayward we can define that {\em a
two-cylinder, $S$, of constant $t$ and $r$ is trapped, marginally trapped,
or untrapped, according to whether $\theta_{+}\theta_{-} > 0$,
$\; \theta_{+}\theta_{-} = 0$, or $\theta_{+}\theta_{-} < 0$.
An apparent horizon, or trapping horizon in Hayward's terminology \cite{Hay94},
is defined as a hypersurface foliated by marginally trapped
surfaces}. It is said {\em outer, degenerate, or inner}, according to
whether $\left.{\cal{L}}_{n}\theta_{+}\right|_{\Sigma} < 0$,
$\left.{\cal{L}}_{n}\theta_{+}\right|_{\Sigma} = 0$, or
$\left.{\cal{L}}_{n}\theta_{+}\right|_{\Sigma} > 0$, where ${\cal{L}}_{n}$
denotes the Lie derivative along the normal direction
${n}_{\mu}$. In addition, if $\left. \theta_{-}\right|_{\Sigma} < 0$
then the apparent horizon is said {\em future}, and if
$\left. \theta_{-}\right|_{\Sigma} > 0$ it is said {\em past} \cite{Hay94}.

On the other hand, from Eq.(\ref{2.7}) we find that
\bq
\lb{2.8}
\alpha_{,\nu}\alpha^{,\nu} = - 2\alpha^{2}e^{2\sigma} \theta_{+}\theta_{-},
\eq
where  $\sigma  \equiv \gamma - \psi +\frac{1}{2} \ln\left(2a'(v)b'(u)\right)$
and is finite on non-singular surfaces \footnote{The functions
$a(v)$ and $b(u)$ can be always chosen 
such that the metric in the $(u,v)$-coordinates 
is free of coordinate singularities on hypersurfaces $u = Const.$ or
$v = Const.$ This implies that $\sigma$ is always finite on these
hypersurfaces, except for the ones on which the spacetime is singular.}. 
Then, from Eq.(\ref{2.8})
we can see that the above definition given in terms of the expansions,
$\theta_{+}$ and $\theta_{-}$, is consistent with that given by Hayward
who defined it according to the nature of the vector $\alpha_{,\mu}$
\cite{Hay00}. This definition is the second criterion used in \cite{Gon02}.

Note that Eq.(\ref{2.7}) is valid for any spacetime where the metric is given by
Eq.(\ref{2.1}). This, in particular,  includes the cases where the spacetimes 
are not vacuum. Thus, to
compare Eq.(\ref{2.2}) with Eq.(\ref{2.7}) we need first to restrict ourselve to
the vacumm case and then choose   the Einstein-Rosen gauge (\ref{2.16a}), 
for which Eq.(\ref{2.7}) reduces to
\bqn
\lb{2.7a}
\theta_{+}  &=& f(u) \frac{1}{r}e^{2(\psi - \gamma)},\nb\\
\theta_{-} &=& - g(v) \frac{1}{r}e^{2(\psi - \gamma)},\;\; (\alpha = r),
\eqn
where $f(u) = 1/(2b'(u))$ and $g(v) = 1/(2a'(v))$. Comparing Eq.(\ref{2.2}) with
Eq.(\ref{2.7a}) we can see that they are quite different, even after the irrelevant
conformal factors $f(u)$ and $g(v)$ are taken into account. Thus, Eq.(\ref{2.2})
must be wrong and all the analysis based on it is incorrect. 

In addition,  the derivation of Eq.(3.8) in \cite{Gon02}
is also wrong, because $t$ and $r$ are no longer independent variables
once Eq.(3.7) holds. This can also be seen as follows: If Eq.(3.8) were correct,
so were Eq.(3.9). Then, from the latter we find  $\psi_{,tr} = 2r^{-2}$ 
and $\psi_{,rt} = 0$, that is, $\psi_{,tr} \not= \psi_{,rt}$, which is
clearly wrong. Here Eqs.(3.7), (3.8) and (3.9) are all referred to
the ones given in \cite{Gon02}. 

From Eq.(\ref{2.7a}), on the other hand, we find
\bq
\lb{2.8a}
\theta_{+}\theta_{-} = - \frac{e^{4(\gamma - \psi)}}{4a'(v)b'(u)r^{2}},
\;\; (\alpha = r),
\eq
which  is always negative, as longer as the quantity $(\gamma - \psi)$ is finite.
This explains why Gon\c{c}alves found no trapped surfaces in the Einstein-Rosen 
coordinates. 

However,  
there exist cases where $(\gamma - \psi)$ becomes unbounded on a hypersurface 
and the singularity on this surface is only a coordinate one. Then, to have a 
geodesically maximal spacetime, extensions beyond this surface are needed.  
After extending the solution to other regions, trapped surfaces may exist. 
To show that this possibility indeed exists, let us consider the following 
solution,
\bqn
\lb{2.9}
\gamma &=&\frac{2n-1}{4n}\ln\left|\frac{f^{4}}{t^{2} - r^{2}}\right|
           + \gamma_{0},\nb\\
\psi&=& q\ln\left|\frac{f}{\sqrt{2}}\right| + \psi_{0},\nb\\
\alpha &=& r,\;\;\;
f = (r - t)^{1/2} + (-t - r)^{1/2},
\eqn
where $\gamma_{0}$ and $\psi_{0}$   are arbitrary constants, and
\bq
\lb{2.10}
q \equiv \left(\frac{2n-1}{n}\right)^{1/2},
\eq
with $n$ being a positive integer, $n \ge 1$.  It can be shown that the
above solutions  satisfy the vacuum Einstein field equations
$R_{\mu\nu} = 0$. On the other hand, from Eq.(\ref{2.9}) we find
\bq
\lb{2.11}
X \equiv \xi^{\alpha}_{(3)}\xi^{\beta}_{(3)}g_{\alpha\beta} = r^{2}
e^{2\psi_{0}}\left|\frac{f}{\sqrt{2}}\right|^{2q}.
\eq
Then, it can be shown that  the local-flatness condition (\ref{cd1}) is
satisfied on the symmetry axis $r = 0$, provided that
$\gamma_{0} = -((2n-1)/n)\ln(2)$. From Eq.(\ref{2.9}) we can see that the
solutions are valid only in the region $0 \le r < - t$. The metric
coefficient $\gamma$ is singular on the hypersurface $r = -t$
($ \gamma \rightarrow - \infty$, as $ r \rightarrow - t$) [cf. Fig. 1].
However, this is
only a coordinate singularity. To see this, let us choose the functions
$a(v)$ and $b(u)$ in Eq.(\ref{2.3a}) as
\bq
\lb{2.11a}
a(v) = - (-v)^{2n}, \;\;\; b(u) = - (-u)^{2n},
\eq
we find that in terms of $u$ and $v$ the metric takes the form 
\bq
\lb{2.1a}
ds^{2} = - 2e^{2\sigma}dudv
+ e^{2\psi}dz^{2} + \alpha^{2} e^{-2\psi}d\varphi^{2},
\eq
with
\bqn
\lb{2.9a}
\sigma &=& - q(1-q)\ln\left|\frac{f}{\sqrt{2}}\right| + \sigma_{0},\nb\\
\psi&=& q\ln\left|\frac{f}{\sqrt{2}}\right| + \psi_{0},\nb\\
\alpha &=& r = (-u)^{2n} - (-v)^{2n},\;\;\;
f = (-u)^{n} + (-v)^{n},
\eqn
where $\sigma_{0} \equiv \frac{1}{2}\ln\left(n^{2}2^{(2-n)/n}\right) - \psi_{0}$.
From Eqs.(\ref{2.3a}) and (\ref{2.11a}) we can see that the symmetry axis $r = 0$
is mapped to $u = v$, and the region $0 \le r < -t$ to the one where $0 > u \ge v$,
which is referred to as  Region $II$ in Fig. 1. The hypersurface $r = -t$ is mapped
to $v = 0$, on which the metric coefficients are no longer
singularity in the ($u,\; v$)-coordinates. This shows clearly that the singularity on the
hypersurface $r = -t$ in the ($t, \; r$)-coordinates is indeed a coordinate one.
Region $I$, where $v > 0, \; u < 0$ and $ |u| > v$, is absent in the $(t,r)$-coordinates,
and such represents an extended region. Across the hypersurface $v = 0$ the metric
coefficients are analytical, and thus the extension is unique.

 \begin{figure}[htbp]
 \begin{center}
 \label{fig1}
 \leavevmode
  \epsfig{file=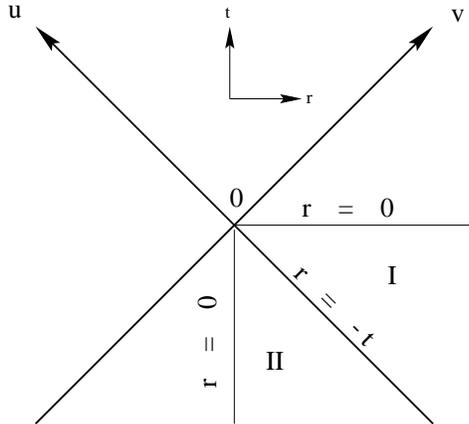,width=0.35\textwidth,angle=0}
 \caption{The spacetime in the $(u, v)$-plane for the solutions given by
 Eqs.(\ref{2.1a}) and (\ref{2.9a}) in the text. Regions $I$ and $II$ are defined,
 respectively, as  $I = \{x^{\alpha}: u < 0,\; v > 0, \; |u| > v\}$, and
 $II = \{x^{\alpha}: u < 0,\; v < 0, \; |u| > |v|\}$. The spacetime is
 locally flat on the vertical line $r = 0$, which is the symmetry axis
 of the spacetime. Region $I$ is
 absent in the $(t,r)$-coordinates.}
 \end{center}
 \end{figure}

On the other hand, from Eqs.(\ref{2.7}) and (\ref{2.9a})  we find
that
\bqn
\lb{2.14}
\theta_{+} &=& 2ne^{-2\sigma} \frac{(-v)^{2n-1}}{r},\nb\\
\theta_{-} &=& -2ne^{-2\sigma} \frac{(-u)^{2n-1}}{r},
\eqn
from which we can see that in Region $II$, where $u, \; v < 0,\; v > u$,
 we have $\theta_{+} > 0$ and $\theta_{-} < 0$. Thus, in this region
the two-cylinders of constant $u$ and $v$ are untrapped. However, 
in Region $I$ we have $\theta_{+}\theta_{-} > 0$ and the
two-cylinders of constant $u$ and $v$ become trapped. 
On the hypersurface $v = 0$ we have
$\theta_{+}(u, 0) = 0$ and $\theta_{-}(u, 0) < 0$. Thus, this hypersurface
defines a future apparent horizon. This horizon is degenerate, 
since now we have $\left.{\cal{L}}{_n}\theta_{+}\right|_{v = 0} = 0$.
The hypersurface $v = -u$ on which we also have
$r = 0$ serves as the up boundary of the spacetime. The nature of the spacetime
singularity on this surface depends on the values of $n$. It can be shown that
when $n$ is an odd integer, the spacetime has curvature singularity there,
and when $n$ is an even integer, the spacetime is free of curvature singularity,
but the local-flatness condition (\ref{cd1}) does not hold. That is, in the
latter case the spacetime has topological singularity at $u = -v$.
The global structure can be seen from the corresponding  ``Penrose diagram,"
Fig. 2, where the quotation means that it is only schematic.

 \begin{figure}[htbp]
 \begin{center}
 \label{fig2}
 \leavevmode
  \epsfig{file=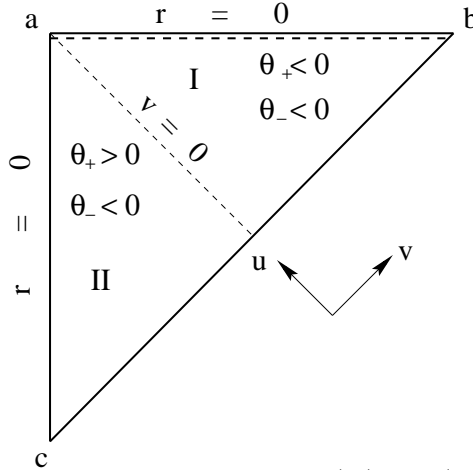,width=0.35\textwidth,angle=0}
 \caption{The ``Penrose diagram" for the solutions given  by
 Eqs. (\ref{2.1a}) and (\ref{2.9a}) with $n$ being a positive integer.
 The two-cylinders of constant $u$ and $v$
are trapped in Region $I$, but not in Region $II$.
 The dashed
line $v = 0$ represents a future degenerate apparent horizon.
When $n$ is an odd integer the spacetime has curvature singularity on the horizontal
line $r = 0$, and when $n$ is an even integer it has no curvature
singularity but a topological one. The line $bc$ represents the past null infinity
$u = -\infty$, and the point $a$ has the coordinates $(u, v) = (0, 0)$.}
 \end{center}
 \end{figure}

The above example  shows clearly  that the Einstein-Rosen coordinates
cover only  a part of the spacetime, region $II$, in which we have
$\theta_{+}\theta_{-} < 0$, that is the two-cylinders of constant $t$
and $r$ are untrapped in this region. However, after the solution is extended to the
whole spacetime, trapped surfaces indeed exist in the extended region,
$I$.  This is quite  similar to the Schwarzschild solution in the spherically
symmetric spacetimes.

\section{Conclusions}

In this Comment, we have shown that Eq.(\ref{2.2}) used by Gon\c{c}alves in
his first criterion for the existence of trapped surfaces in the cylindrical
spacetimes is incorrect \cite{Gon02}. We have also shown that the Einstein-Rosen
coordinates used by Gon\c{c}alves do not always
cover the whole spacetime. As a result, his proof that trapped surfaces
don't exist in vacuum is incomplete.  In fact, we have presented an example,
which is a  solution to the Einstein vacuum field equations
and satisfies all the conditions imposed by  Gon\c{c}alves. After extending the
solution to the whole spacetime we have shown explicitly that  trapped surfaces
exist in the extended region.

It should be noted that lately Gon\c{c}alves generalized his studies to the
case where the cylindrical gravitational waves have two degrees of polarization
\cite{Gon03}. Following the arguments given above, one can show that the
coordinates used there do not always cover the whole spacetime either, and
the analysis for the non-existence of trapped surfaces in the vacuum part of
the spacetime is also incomplete.

We would also like to  note that the results obtained here do not contradict
with the ones obtained by Berger, Chrusciel and Moncrief \cite{BCM95},
since in the present case it can be shown that the spacetime is not
``asymptotically flat" in the sense defined by them. In addition, 
our results do not contradict with the ones obtained by Ida either \cite{Ida00}. The reason is simply because Ida defined black hole in terms 
of {outer} apparent horizons, while the apparent horizons appearing in the 
present solutions are degenerate.

\section*{Acknowledgments}

The author  thanks  Sean A. Hayward for carefully reading the manuscript and
valuable discussions and comments. He would also like to thank  Y.T. Liu for
delighted conversations and comments, and  the Department of Physics, UIUC, for
hospitality.

\end{document}